\documentclass[letterpaper, 10 pt, conference]{ieeeconf}

\IEEEoverridecommandlockouts                 
\usepackage{cite}
\usepackage{amssymb,amsfonts}
\usepackage{algorithmic}
\usepackage{graphicx}
\usepackage{textcomp}
\usepackage{xcolor}
\usepackage{algorithm}
\usepackage{xcolor,cancel}
\usepackage{dsfont}
\usepackage{amsmath}
\usepackage{soul,xcolor}
\usepackage{stmaryrd}
\usepackage{balance}
\usepackage{url}

\newcommand{\cD}{{\cal D}}

\newcommand{\cN}{{\cal N}}

\newcommand{\cG}{{\cal G}}
\newcommand{\cA}{{\cal A}}
\newcommand{\cS}{{\cal S}}
\newcommand{\cR}{{\cal R}}

\newcommand{\cI}{{\cal I}}

\newcommand{\cX}{{\cal X}}

\newcommand{\sF}{\textsf{F}}

\newcommand{\sfN}{\textsf{N}}

\newtheorem{theorem}{Theorem}

\newtheorem{proposition}{Proposition}

\newtheorem{definition}{Definition}
\newtheorem{assumption}{Assumption}
\newtheorem{problem}{Problem}
\newtheorem{remark}{Remark}

\newcommand{\real}{\mathbb{R}}

\allowdisplaybreaks[3]

\DeclareMathAlphabet{\mymathbb}{U}{BOONDOX-ds}{m}{n}

\newcommand{\zero}{\mymathbb{0}} 

\def\BibTeX{{\rm B\kern-.05em{\sc i\kern-.025em b}\kern-.08em
    T\kern-.1667em\lower.7ex\hbox{E}\kern-.125emX}} 
\title{Local Safety Filters for Networked Systems \\ via Two-Time-Scale Design}

\author{Emiliano Dall'Anese
\thanks{E. Dall'Anese is with the Department of Electrical and Computer Engineering and the Division of Systems Engineering, Boston University,
Boston, MA 02215, USA (email: edallane@bu.edu). This work was supported in part by the AFOSR Award FA9550-23-1-0740.}}

\begin{document}

\maketitle

\begin{abstract}
Safety filters based on Control Barrier Functions (CBFs) provide formal guarantees of forward invariance, but are often difficult to implement in networked dynamical systems. This is due to global coupling and communication requirements. 
This paper develops locally implementable approximations of networked CBF safety filters that require no coordination across subsystems. 
The proposed approach is based on  a two-time-scale dynamic implementation inspired by singular perturbation theory, where a small parameter $\epsilon$ separates fast filter dynamics from the plant dynamics; then, a local implementation is enabled via derivative estimation. 
Explicit bounds are derived to quantify the mismatch between trajectories of the systems with dynamic filter and with the ideal centralized safety filter. 
These results characterize how safety degradation depends on the time-scale parameter $\epsilon$, estimation errors, and filter activation time, thereby quantifying trade-offs between safety guarantees and local implementability.
\end{abstract}


\section{Introduction}
\label{sec:introduction}

Control Barrier Functions (CBFs) provide a framework for enforcing forward invariance of prescribed safe sets~\cite{ames2017control}. 
In this context, CBFs are commonly used to construct safety filters that minimally modify the input of a nominal controller, which are designed for stability or robustness, to enforce safety. 

In networked dynamical systems, such as transportation networks~\cite{cassandras2025control}, power grids~\cite{dorfler2023control}, and robotic networks~\cite{wabersich2023data}, safety poses additional challenges: constraints are often defined locally, while the underlying dynamics are globally coupled. 
In these settings, enforcing safety via CBF-based filters generally requires solving a network-level optimization problem that captures interconnection effects. 
Such implementations can be difficult to realize in practice due to communication requirements and  control update rates. 
This gap between theoretical safety guarantees and implementable architectures motivates the study of distributed and locally implementable safety filters and the characterization of trade-offs between safety, communication, and the availability of model information.

\emph{Prior works}. Distributed implementations of CBF-based safety filters have been studied by decomposing centralized CBF quadratic programs across agents~\cite{tan2021distributed,wang2025distributed}. 
Related approaches have developed distributed CBF-based controllers for safe navigation and formation control~\cite{mestres2024distributed}, where coupling is through  the safety constraints. 
Alternative constructions of CBFs have been proposed to improve implementability by reducing computational complexity~\cite{min2025tac}. 
Scalable formulations based on Banach CBFs for continuum swarm models have also been explored~\cite{gao2026banach}. 
Collaborative approaches have been proposed for dynamically coupled networks, where neighboring agents coordinate safety actions through communication~\cite{butler2025collaborative}. 
Collectively, these works primarily focus on distributed implementations (with different degrees of communication overhead) of safety filters. 
Open questions remain on how to realize fully local safety filters with minimal communication and limited model information.

\emph{Contributions}. This paper studies safety filters that can be implemented locally at each subsystem. 
The main contributions are summarized as follows.

\noindent \emph{(c1)} The paper considers a class of CBF-QPs with separable structure that admit closed-form safety filters. Since these filters still require global state information and system dynamics, the paper develops locally implementable approximations using a two-step design. First, inspired by singular perturbation theory~\cite{kokotovic1976singular}, a dynamical approximation is introduced, in which a small parameter $\epsilon$ induces a separation between fast filter dynamics and plant dynamics. Second, the paper leverages local derivative estimates (e.g., dirty derivatives~\cite{astrom1995pid,marchi2022dirty}) to obtain dynamic safety filters that rely only on local information.

\noindent \emph{(c2)} The paper derives explicit trajectory deviation bounds between the systems equipped with dynamic filters and centralized safety filters with network and disturbance information. The bounds characterize how safety degradation depends on the time-scale parameter $\epsilon$, estimation errors of the derivative, and the availability of state and network model, thereby quantifying trade-offs between safety guarantees and implementability. 
To the best of the author's knowledge, this  work provides a first trajectory-level characterization of safety degradation in locally implementable CBF filters in network systems. 

\noindent \emph{(c3)} The proposed dynamic filters is evaluated on frequency control in power transmission systems with grid-following inverters~\cite{dorfler2023control}. Constraints on the frequency nadir following generation losses or load increases are imposed. The results illustrate how safety degradation scales with  $\epsilon$ while demonstrating the feasibility of fully local implementations.

Lastly, two-time-scale proximal primal–dual flows for tracking centralized CBF-QP solutions were proposed in~\cite{marvi2024control}. This paper develops locally implementable approximations of centralized CBF filters with explicit trajectory-level guarantees.

\section{Preliminaries and system model}
\label{sec:model}

\emph{Notation}. The set of real numbers is denoted by $\real$.
Given vectors $x\in \real^n$ and $y\in \real^m$, let $(x,y)\in \real^{n+m}$ denote their concatenation; given matrices $A \in \real^{n\times n}$, $B \in \real^{m\times m}$, $\mathrm{blkdiag}(A,B)$ forms a block-diagonal matrix. Given a norm $\|\cdot\|_x$ on $\real^n$, $\|x\|_x$ denotes the norm of $x\in\real^n$, and $\|A\|_x$ denotes the induced matrix norm of $A\in\real^{n\times n}$.
The induced logarithmic norm (log-norm) associated with $\|\cdot\|_x$ is defined as  $\mu_{\|\cdot\|_x}(M)
:=\lim_{h\to 0^+}\frac{\|I_n+hM\|_x-1}{h}$. Given norms $\|\cdot\|_x$ on $\real^n$ and $\|\cdot\|_z$ on $\real^m$, and a matrix $M\in\real^{m\times n}$, the induced cross-norm is $\|M\|_{x\to z}:=\max_{\|v\|_x=1}\|Mv\|_z$.
Finally, $\llbracket \cdot \,;\, \cdot \rrbracket_x$ denotes a compatible weak pairing  associated with the norm $\|\cdot\|_x$~\cite[Ch.~2]{FB:24-CTDS}. To simplify notation, the paper drops  subscripts; the  norms will be understood from context.

A continuous function  $\alpha:\real\to\real$ is an extended class  $\mathcal{K}_{\infty}$  if $\alpha(0)=0$,
$\alpha$ is strictly increasing and $\lim\limits_{s\to\pm\infty}\alpha(s) = \pm\infty$; a natural example is the function $\alpha(x) = \alpha_0 x$, with $\alpha_0 \in \real_{>0}$. For a real-valued function $f:\real\to\real$, the upper right-hand Dini derivative is defined as $D^+ f(t):=\limsup_{h\to 0^+}\frac{f(t+h)-f(t)}{h}$. Given a function $f:\real^n\to\real^m$,  $f$ is Lipschitz on a set $\cX\subseteq\real^n$ with constant $\ell_{f,x}$ (with respect to $\|\cdot\|_x$ and $\|\cdot\|_z$) if $\|f(x)-f(y)\|_z \le \ell_{f,x}\,\|x-y\|_x,
\, \forall x,y\in\cX$. All functions are assumed measurable where needed, and locally integrable.

\subsection{Network model and safety objectives}

Consider a network of $N$ subsystems. For each subsystem $i\in \cN := \{1,\dots,N\}$,
let $x_i\in\real^{n_i}$ denote the local state, $u_i\in\real^{m_i}$ the local input,
and $w_i\in\real^{n_i}$ an unknown exogenous input or disturbance. Let
$x:= (x_1,\dots,x_N)\in\real^{n}$ with $n:=\sum_{i=1}^N n_i$,
$u:= (u_1,\dots,u_N)$, and $w:=(w_1,\dots,w_N)$.
Consider networked dynamics of the form
\begin{equation}\label{eq:network_dynamics}
\dot x_i = f_i(x) + B_i u_i + w_i,\qquad i=1,\dots,N,
\end{equation}
where the map $x\mapsto f_i(x)$ is locally Lipschitz on an open convex
domain $\cD \subset \real^n$ (for a given norm $\|\cdot\|$ on $\real^n$), $B_i \in \real^{n_i \times m_i}$, and $t \mapsto w(t)$ is measurable and essentially bounded. The local states $\{x_i\}_{i\in\cN}$ are coupled through the vector fields $\{f_i(x)\}_{i\in\cN}$,
which capture the network interactions. 
Stacking~\eqref{eq:network_dynamics} 
and defining $f(x):=(f_1(x),\dots,f_N(x))$ and
$B:=\mathrm{blkdiag}\big(B_1,\dots,B_N\big)$ yields
\begin{equation}\label{eq:network_dynamics_stacked}
\dot x = f(x) + B u + w . 
\end{equation}

It is assumed that each subsystem implements a local nominal controller
$u_i = \kappa_i(x_i)$,
where $\kappa_i:\real^{n_i}\to\real^{m_i}$ is locally Lipschitz.
Let $\kappa(x):=(\kappa_1(x_1),\dots,\kappa_N(x_N))$ and $m := \sum_{i=1}^N m_i$; then the
\emph{nominal} dynamics are given by
\begin{equation}\label{eq:network_nominal_cl}
\dot x = \sF(x) + w,
\qquad
\sF(x):= f(x) + B\kappa(x).
\end{equation}
The nominal controllers $\kappa_i$
are designed to achieve stabilization of desired equilibria of the networked system. In particular, the following is assumed. 

\begin{assumption}[Stable nominal dynamics] 
\label{as:F} 
For a given norm $\|\cdot\|$ on $\real^n$, there exists $c_\sF < 0$ such that $\mu(D \sF (x)) \leq c_\sF$ for almost every $x \in \cD$. \hfill $\Box$ 
\end{assumption} 

Since $x \mapsto \sF(x)$ is locally Lipschitz, it is differentiable almost everywhere by Rademacher’s theorem. Assumption~\ref{as:F} implies that, for any given $w$,  the nominal system is strongly infinitesimally contractive in the norm $\|\cdot\|$~\cite{davydov2025tac}, hence incrementally exponentially stable on $\mathcal D$ ($c_\sF$ is essentially the one-sided Lipschitz constant of $\sF(x)$).

The goal is to enforce safety constraints on the local states.
Specifically, for each subsystem $i$, let $\cS_i \subset \mathbb{R}^{n_i}$  denote
a local \emph{safe set}. Given a continuously differentiable function
$h_i:\cD\to\real$, the local safe set  is defined as
\begin{equation}\label{eq:local_safe_set}
\mathcal S_i := \{x_i\in\real^{n_i}: h_i(x_i)\ge 0\}.
\end{equation}
The overall safe set for the network is given by
$\cS:=\{x\in\cD:\ h_i(x_i)\ge 0,\ \forall i\in\cN\}$.
Since the nominal controllers are not guaranteed to ensure safety,
a common approach is to provide an input correction by leveraging tools from  CBFs~\cite{ames2017control}. In the following, the
definition of CBFs is recalled (with definitions provided globally for the network).

\begin{definition}[Network CBF]
\label{defi:cbfnetwork}
The functions $\{h_i\}_{i\in\cN}$ define a CBF for $\cS$ on $\cD$ under the dynamics~\eqref{eq:network_nominal_cl}
if there exist extended class-$\mathcal K_\infty$ functions
$\alpha_i$ such that, for all $x\in\cD$, there exist inputs $u_i$
satisfying $\nabla h_i(x_i)^\top \big(\sF_i(x) + B_iu_i + w_i\big)
+ \alpha_i(h_i(x_i)) \ge 0$ for all $i\in\cN$. \hfill $\Box$
\end{definition}

To minimally modify the local nominal controllers, introduce a correction $s \in\mathbb R^{m}$ and set $u_i = \kappa_i(x_i) + s_i$, $i \in \cN$. In particular, for a given state $x$, the correction $s(x)$, also referred to as \emph{safety filter}, is computed as the unique optimal solution of the following quadratic program (QP)~\cite{ames2017control}: 

\vspace{-.4cm}

\begin{subequations}
\label{eq:qp}
\begin{align}
    s(x) := \arg & \min_{\theta \in \real^m} \|\theta\|_2^2 \label{eq:qp-cost} \\
    & \text{subject to}:~ \nabla h_i(x_i)^\top \big(\sF_i(x) + B_i\theta_i + w_i\big) ~~~~~~~~~~ \nonumber \\
    & ~~~~~~~~~~~~~~~~ + \alpha_i(h_i(x_i)) \ge 0, \,\,\,\, i \in \cN \label{eq:qp-constraints}
\end{align}
\end{subequations}
where $\theta = (\theta_1,\dots,\theta_N)$ and $\sF_i(x) = f_i(x) + B_i \kappa_i(x_i)$.  The following assumption is imposed (cf.~\cite{ames2017control}).

\begin{assumption}[Well-posedness of the safety filter]
\label{as:static_wellposed}
For every $x\in\cD$, the QP~\eqref{eq:qp} is feasible. For all $i\in\cN$ and all $x$ such that $h_i(x_i)=0$, it holds that  $B_i^\top \nabla h_i(x_i) \neq 0$.
\hfill $\Box$
\end{assumption}

The resulting \emph{filtered} system is then defined as
\begin{align}
\label{eq:filtered_system}
\dot x = \sfN(x) := \sF(x) + B s(x) + w \, .
\end{align}
Under Assumption~\ref{as:static_wellposed}, $s(x)$ is well defined on $\cD$; moreover,  the right-hand side of~\eqref{eq:filtered_system}
is locally Lipschitz~\cite{mestres2025regularity}. When $\{h_i\}_{i\in\cN}$ define a CBF for $\cS$ on $\cD$ as in Definition~\ref{defi:cbfnetwork}, it is well known that~\eqref{eq:filtered_system} 
renders the set $\cS$ forward invariant;
see, e.g.,~\cite{ames2017control} for a formal proof of this result.

\subsection{Problem statement}

Although each constraint in the optimization problem~\eqref{eq:qp} depends on the global state $x$ through $\sF_i(x)$, the optimization variables are decoupled, since the $i$th constraint depends only on $\theta_i$. Therefore,~\eqref{eq:qp} decouples into $N$ local subproblems, one per subsystem $i \in \cN$, of the form
\begin{align*}
& \min_{\theta_i \in \real^{m_i}} \|\theta_i\|_2^2 \\
& ~ \text{subject to:}
~ \nabla h_i(x_i)^\top \big(\sF_i(x) + B_i\theta_i + w_i\big) + \alpha_i(h_i(x_i)) \ge 0.
\end{align*}
For fixed $x$, this problem corresponds to the projection of $\zero_{m_i}$ onto a half-space~\cite[Ch.~8]{boyd2004convex}, and it admits the solution
\begin{align}
    \label{eq:qp_solutionReLU}
    s_i(x) = d_i(x_i)\max\{0,-\eta_i(x)\}
\end{align}
where
\begin{align*}
\eta_i(x) &:= \nabla h_i(x_i)^\top \big(\sF_i(x) + w_i\big) + \alpha_i(h_i(x_i)), \\
d_i(x_i) &:= \frac{B_i^\top \nabla h_i(x_i)}{\|B_i^\top \nabla h_i(x_i)\|^2}
\end{align*}
for brevity. Computing $s_i(x)$ at each subsystem $i\in\cN$ requires access to:
(i) the model terms $\sF_i(x)$, which depend on the global state $x$ and the network couplings,
and (ii) the exogenous input $w_i$.
In many large-scale systems, such as
power grids~\cite{dorfler2023control} transportation systems~\cite{cassandras2025control}, and robotic networks~\cite{wabersich2023data}, 
collecting the global state in real time is impractical due to communication and latency constraints.
As a result, evaluating $\sF_i(x)$ at the time scales required for safety enforcement
may be impractical. Furthermore, disturbance observers become difficult to design
when accurate models of $\sF_i(x)$ are unavailable.
This fundamentally limits the deployment of CBF-based safety filters
in distributed, real-time settings and motivates the main problem tackled in this paper.

\vspace{.1cm}

\begin{problem} [Filter design]
\label{prob:design}
Design an approximate safety filter that can be implemented locally at each
subsystem without requiring knowledge of the global state $x$,  function $\sF(x)$, and the disturbance $w_i$.
Characterize the trade-offs between safety and the availability
of model and state information. \hfill $\Box$
\end{problem}

\vspace{.1cm}

A couple of remarks are provided next.

\vspace{.1cm}

\begin{remark}[Safety sets]
\label{rem:closed_form}
For clarity of exposition, the proposed approach is presented for local safe
sets defined as in~\eqref{eq:local_safe_set}. However, the methodology developed
in this paper is applicable to cases when $\cS_i$ is defined by multiple constraints; the key requirement
is that the safety filter~\eqref{eq:qp} admits a closed-form solution. This condition holds, for example, when enforcing box constraints on the local states $x_i$, when the matrix of  the linear constraints is order preserving, or when log-sum-exp smoothing is used.  See, for example~\cite{mestres2025explicit}.\hfill $\Box$
\end{remark}

\vspace{.1cm}

\begin{remark}[System dynamics]
\label{rem:system_dynamics}
The technical arguments of the paper can be, at the cost of a  heavier notation, be extended to dynamics of the form  $\dot x_i = f_i(x) + g_i(x_i) u_i + w_i$, with $g_i(x_i)$ locally Lipschitz on $\cD$. \hfill $\Box$
\end{remark}

\section{Local Dynamic Filter}
\label{sec:dynamic_filter}


\subsection{Design of the Dynamic Filter}
\label{sec:dynamic_filter_feedback_design}

The main  impediment to local evaluation of the filter $s_i(x)$
in~\eqref{eq:qp_solutionReLU} is the term
$\nabla h_i(x_i)^\top \big( \sF_i(x) + w_i \big)$
appearing in $\eta_i(x)$. To eliminate the need for $\sF_i(x)$ and $w_i$, this paper constructs an
approximate dynamic filter via a two-step procedure.

\noindent \emph{(i) Dynamic relaxation}. First, the proposed approach considers a dynamical approximation of~\eqref{eq:qp_solutionReLU} 
\begin{align}
\label{eq:qp_solutionReLU_dyna_approx}
\dot z_i = - z_i + s_i(x), \,\,\,\, i \in \cN , 
\end{align}
and utilizes $z := (z_1, \ldots, z_N)$ as the input correction to the system; i.e.,  $ \dot{x}= \sF(x) + Bz + w$. For frozen $x$, the dynamics~\eqref{eq:qp_solutionReLU_dyna_approx} are exponentially stable, implying that $z_i$ converges to $s_i(x)$.
This observation will serve as the basis for the two-time-scale
construction~\cite{kokotovic1976singular}.

\noindent \emph{(ii) Measurement-based approximation}. Second, using $z$ as the correction, note that $\dot{x}_i -  B_iz_i = \sF_i(x) + w_i$.  If an estimate $\widehat{\dot x}_i$ of $\dot x_i$ is available,~\eqref{eq:qp_solutionReLU_dyna_approx} can be approximated by
\begin{subequations}
\label{eq:s_hat}
\begin{align}
& \dot z_i = - z_i + d_i(x_i) \max\{0,-\widehat{\eta}_i(x;\widehat{\dot x}_i)\}  \\
& \widehat{\eta}_i(x;\widehat{\dot x}_i):= 
\nabla h_i(x_i)^\top (\widehat{\dot x}_i  - B_iz_i)
\!+ \! \alpha_i(h_i(x_i)).
\end{align}
\end{subequations}
The dynamics~\eqref{eq:s_hat} are locally implementable at subsystem $i$
and require only local derivative estimates. 

Given~\eqref{eq:s_hat}, the following  two-time-scale implementation is proposed:

\vspace{-.4cm}

\begin{subequations}
\label{eq:tts_meas}
\begin{align}
\dot x &= \sF(x) + Bz + w, \label{eq:tts_meas_velocity_x}\\
\epsilon \dot z_i &= - z_i + \tilde{s}_i(x_i,z_i;\widehat{\dot x}_i),
\qquad i \in \cN, \label{eq:tts_meas_z}
\end{align}
\end{subequations}
where
$\tilde{s}_i(x_i,z_i;\widehat{\dot x}_i)
:=  d_i(x_i)\max\{0,-\widehat{\eta}_i(x,\widehat{\dot x}_i)\}$, and $\epsilon>0$ induces a time-scale separation between
the dynamic filter and the plant. In particular, when $\epsilon \ll 1$, the filter dynamics evolve on the
fast time scale~\cite{kokotovic1976singular}. The paper will show that $\epsilon$ tunes the
bandwidth of the dynamic filter and provides a built-in trade-off between safety and noise
attenuation.

Borrowing the terminology from singular-perturbation theory~\cite{kokotovic1976singular},
 the static filtered system is now connected with the
two-time-scale system~\eqref{eq:tts_meas}.
In the ideal case where $\widehat{\dot x}=\dot x$ (i.e., perfect estimate), the fast dynamics
reduce to $\epsilon \dot z = -z + s(x)$.
In the limit $\epsilon\to 0^+$, the slow manifold is given by $z=s(x)$,
and the reduced model coincides with~\eqref{eq:filtered_system}.
Next, inspects the case with estimation errors; 
suppose that
\begin{equation}
\label{eq:error_model}
\widehat{\dot x}_i=\dot x_i +e_{i},
\end{equation}
where $e_i:\,t\to\real^{n_i}$ is measurable and essentially bounded. 
Noting that  $\nabla h_i(x_i)^\top (\widehat{\dot x}_i  - B_iz_i) = \nabla h_i(x_i)^\top (\dot x_i + e_i  - B_iz_i) = \nabla h_i(x_i)^\top (\sF_i(x) + w_i) + \nabla h_i(x_i)^\top e_i$, in the limit $\epsilon\to 0^+$,~\eqref{eq:tts_meas_z}  
yields the correction
\begin{equation}
\label{eq:perturbed_static_filter}
s_{e,i}(x,t)
:=
d_i(x_i)\max\{0,-\eta_i(x)-\nabla h_i(x_i)^\top e_{i}(t)\},
\end{equation}
which can be interpreted as a perturbed version of $s(x)$.

\vspace{.1cm}

\begin{remark}[Estimation of the derivative]
\label{rem:dirty_derivative}
The error model in~\eqref{eq:error_model} is intentionally kept general so as to accommodate different schemes for the derivative estimation. A simple locally implementable example is the ``dirty derivative,'' for which the estimate is generated via $\dot \rho_i = -\frac{1}{\tau_d} \rho_i + \frac{1}{\tau_d} x_i$, 
$\widehat{\dot x}_i = \frac{1}{\tau_d}(x_i-\rho_i)$
where $\tau_d>0$ is a parameter~\cite{astrom1995pid,marchi2022dirty}. On finite intervals where the relevant closed-loop trajectories remain bounded, this estimator is consistent with~\eqref{eq:error_model}, since it produces an estimate $\widehat{\dot x}_i$ and a measurable, essentially bounded error $e_i=\widehat{\dot x}_i-\dot x_i$. Observer-based schemes or disturbance observers may provide improved estimates of $\dot x_i$, but they typically require additional model information and may be less suitable for fully local implementations.  \hfill $\Box$
\end{remark}

\subsection{Analysis of the Dynamic Filter}
\label{sec:dynamic_filter_analysis_trajectory}

The analysis is based on bounding the distance between trajectories of~\eqref{eq:tts_meas} and those of the ideal filtered system~\eqref{eq:filtered_system}. This viewpoint is natural because~\eqref{eq:filtered_system} guarantees forward invariance of the safe set. By quantifying the trajectory deviation induced by the dynamic implementation,
one can obtain explicit trade-offs between time-scale separation, estimation accuracy,
and safety degradation. In what follows, $\ell_{s,x}$ denotes the Lipschitz constant of $s(x)$ on $\cD$, while 
$\ell_{s,e}:=\sup_{x\in \cD}\|d(x)\nabla h(x)^\top\|$ ($d(x):=\mathrm{blkdiag}(d_1(x_1),...,d_N(x_N))$). All the results are for given vector norms and corresponding induced norms.  

\vspace{.1cm}

\begin{proposition}[Error of the dynamic filter]
\label{lem:z_tracks_s_bound}
Consider the dynamics~\eqref{eq:tts_meas} over  $[t_0,t_1]$. Suppose that
$x(t)\in \cX \subset\cD$ for all $t\in[t_0,t_1]$, with $\cX$ compact, and let Assumption~\ref{as:static_wellposed} hold. Define $\tilde{z} := z(t)-s(x(t))$. If $\epsilon^{-1} > \ell_{s,x}\|B\|$, then, 
\begin{equation}
\label{eq:z_bound}
\|\tilde{z}(t)\| \le e^{-(\frac{1}{\epsilon}-\ell_{s,x}\|B\|)(t-t_0)} \|\tilde{z}(t_0)\|
+ E(\epsilon,t_1)
\end{equation}
for all $t\in[t_0,t_1]$, where $$
E(\epsilon,t_1):=\frac{\epsilon\,\ell_{s,x}\,\bar N(t_1)+\ell_{s,e}\,\bar e(t_1)}{1-\epsilon\,\ell_{s,x}\|B\|}$$
with $\bar e(t_1):=\operatorname*{ess\,sup}_{\tau\in[t_0,t_1]}\|e(\tau)\|$, and $\bar N(t_1)$ defined as $\bar N(t_1):=\sup_{\tau\in[t_0,t_1]}\|\sfN(x(\tau))\|$. \hfill $\Box$
\end{proposition}

\vspace{.1cm}

The proof is postponed to Section~\ref{sec:proofs}. The bound~\eqref{eq:z_bound} reveals a fundamental trade-off.
The parameter $\epsilon$ acts as a bandwidth knob:
as $\epsilon \to 0^+$, the dynamic filter approaches the static
 safety filter, with vanishing  tracking error in the noise-free case; i.e., $\|\tilde{z}(t)\| \rightarrow 0$ when $\bar e(t_1) = 0$. However, when derivative estimates have errors, the bias is lower bounded by $\mathcal O(\ell_{s,e}\bar e(t_1))$, highlighting an intrinsic robustness–responsiveness trade-off. The restriction to a compact set $\mathcal{X}$ is standard and can be ensured by boundedness of the closed-loop trajectories under the dynamic filter~\cite{Khalil}.

\vspace{.1cm}

\begin{theorem}[{Distance between $x(t)$ and $x_s(t)$}]
\label{thm:traj_deviation}
Consider the solution $(x(t),z(t))$ of the two-time-scale filtered  system~\eqref{eq:tts_meas} starting from $(x(t_0),z(t_0))$, and let $x_s(t)$ be the solution of~\eqref{eq:filtered_system} starting from $x_s(t_0)$. Assume that: $x(t),x_s(t)\in\cX\subset\cD$ for all $t\in[t_0,t_1]$, with $\cX$ compact; Assumptions~\ref{as:F}--\ref{as:static_wellposed} hold; and, $\epsilon^{-1}-\ell_{s,x}\|B\|>0$.
Define the inactive time set
\[
\cI:=\{t\in[t_0,t_1]:\ s(x(t))=0 \ \text{and}\ s(x_s(t))=0\},
\]
and let $\cA:=[t_0,t_1]\setminus\cI$. Then, for all $t\in[t_0,t_1]$, the following holds for $\tilde x(t) := x(t) - x_s(t)$:
\begin{align}
\|\tilde x(t)\|
\le & 
e^{c_\sF(t-t_0)+\ell_{s,x}\|B\|T_{\cA}(t)}\,\|\tilde x(t_0)\| \label{eq:tilde_x_simplified_bound} \\
& \hspace{-.4cm} +\|B\|\,e^{\ell_{s,x} \|B\|\,T_{\cA}(t)}
\left[\!
\frac{\|\tilde z(t_0)\|}{\epsilon^{-1}\!-\!\ell_{s,x}\|B\|}
+\frac{1}{|c_\sF|}\,E(\epsilon,t_1)
\! \right] \nonumber 
\end{align}
where $T_{\cA}(t):=\int_{t_0}^{t}\mathbf 1_{\cA}(\tau)\,d \tau$. 
\hfill $\Box$
\end{theorem}

\vspace{.1cm}

The proof is postponed to Section~\ref{sec:proofs}. Some observations are in order.

\noindent \emph{i)} In the bound~\eqref{eq:tilde_x_simplified_bound}, the parameter $\epsilon$ enters only through the term $E(\epsilon,t_1)$ (see Proposition~\ref{lem:z_tracks_s_bound}).
As $\epsilon\to 0^+$, the deviation $\|\tilde x(t)\|$ approaches an $\epsilon$-independent limit whenever the estimation error $\bar e(t_1)$ does not vanish.
If $\bar e(t_1)=0$, the dominant contribution becomes $E(\epsilon,t_1)=O(\epsilon)$.

\noindent \emph{ii)} A more negative $c_\sF$ strengthens the contraction of the nominal closed-loop system and reduces the contribution of the error term proportional to $1/|c_\sF|$. However, $c_\sF<0$ alone does not imply decay of the first term in~\eqref{eq:tilde_x_simplified_bound}, since this also depends on the activation term $\|B\|\ell_{s,x}T_{\cA}(t)$ (note that this term vanishes whenever $\|\tilde x(t_0)\|=0$).

\noindent \emph{iii)} The quantity $c_\sF(t-t_0)+\|B\|\ell_{s,x}T_{\cA}(t)$ captures the competition between nominal contraction and filter activation. Here, $c_\sF(t-t_0)$ reflects the baseline contraction of the nominal dynamics, whereas $\|B\|\ell_{s,x}T_{\cA}(t)$ accumulates only over intervals in which both safety filters are active. Thus, if the filter is active rarely, so that $T_{\cA}(t)\ll t-t_0$, the contraction induced by $c_\sF$ can dominate; if instead the filter is active frequently, then $T_{\cA}(t)$ may scale as $t-t_0$, and no domination can be concluded without additional conditions.

The dynamic filter~\eqref{eq:tts_meas} does not, in general, guarantee safety. Nevertheless, the bound~\eqref{eq:tilde_x_simplified_bound} quantifies how far the trajectory $x(t)$ may deviate from the reference trajectory $x_s(t)$ generated by the static safety filter, thereby providing an estimate of the potential safety violation as a function of $\epsilon$ and the estimation error. This reflects an inherent tradeoff: exact safety can be enforced by the ideal static filter using full model and global state information, whereas the proposed dynamic filter relaxes this requirement in exchange for local implementability. One possible remedy is to use a tightened safe set, and tune $\epsilon$ so that the dynamic filtered trajectory remains in the original safe set.

Finally, while  Theorem~\ref{thm:traj_deviation} parallels classical singular perturbation intuition~\cite{kokotovic1976singular,Khalil}, the bound in~\eqref{eq:tilde_x_simplified_bound} is of a different nature: it is derived directly for the dynamic safety filter and explicitly captures the effects of filter activation time and derivative-estimation errors. Unlike contraction-based singular perturbation results such as~\cite{del2012contraction} and~\cite{cothren2023online}, the analysis does not require contraction of the fast and slow subsystems or contraction of the reduced model.

\section{Proofs}
\label{sec:proofs}

\subsection{Proof of Proposition~\ref{lem:z_tracks_s_bound}}

Recall that $\tilde z(t):=z(t)-s(x(t))$. For brevity, let $G(x,z) = \epsilon^{-1}(- z + s(x))$ and $\tilde{G}(x,z) = \epsilon^{-1}(- z + \tilde{s}(x,z;\widehat{\dot x}))$, so that the ideal fast dynamics read $\dot z = G(x,z)$ and the implemented ones read
$\dot z = \tilde G(x,z)$.  Let $\llbracket \cdot ; \cdot \rrbracket$ be a compatible weak pairing associated with the selected norm $\|\cdot\|$ on $\real^m$~\cite[Ch.~2]{FB:24-CTDS}.

Using the curve norm derivative formula, one has that 
$$
\|\tilde z(t)\| D^+\|\tilde z(t)\| \leq  \llbracket \dot z(t)-D s(x(t))\,\dot x(t) ; \tilde z(t) \rrbracket, 
$$
for a.e. $t \in [t_0,t_1]$, where $D s(x(t))$ is the Jacobian of $s(x)$ evaluated at $x(t)$ (defined for a.e.\ $x$ by Rademacher's theorem). By sub-additivity of the weak pairing and the Cauchy--Schwarz-type bound
$\llbracket v;w\rrbracket \le \|v\|\,\|w\|$ for compatible pairings~\cite[Ch.~2]{FB:24-CTDS}, one obtains:
$$
\|\tilde z(t)\|\,D^+\|\tilde z(t)\|
\le
\llbracket \dot z(t)\,;\,\tilde z(t)\rrbracket
+
\|D s(x(t))\,\dot x(t)\|\,\|\tilde z(t)\|.
$$

First, bound the term $\llbracket \dot z;\tilde z\rrbracket$.
Using $\dot z = \tilde G(x,z)$, add and subtract $G(x,z)$ to obtain $\llbracket \dot z\,;\,\tilde z\rrbracket
=
\llbracket \tilde G(x,z)\,;\,\tilde z\rrbracket
=
\llbracket G(x,z)\,;\,\tilde z\rrbracket
+
\llbracket \tilde G(x,z)-G(x,z)\,;\,\tilde z\rrbracket$. 
Note that $\llbracket G(x,z)\,;\,\tilde z\rrbracket = \llbracket G(x,z) - G(x,s(x))\,;\,\tilde z\rrbracket$; then, from~\cite[Thm.~3.7]{FB:24-CTDS}, the first term satisfies
\begin{equation}
\label{eq:pairing_contraction}
\llbracket G(x(t),z(t))\,;\,\tilde z(t)\rrbracket
\le
-\epsilon^{-1}\|\tilde z(t)\|^2,
\end{equation}
that is, $G(x(t),z(t))$ is strongly infinitesimally contractive in $z$ for any given $x$. For the mismatch term, use $\llbracket v;w\rrbracket\le \|v\|\,\|w\|$ to obtain 
$\llbracket \tilde G(x,z)-G(x,z)\,;\,\tilde z\rrbracket
\le
\|\tilde G(x,z)-G(x,z)\|\,\|\tilde z\|$. 
Finally, by the definitions of $G$ and $\tilde G$, $\tilde G(x,z)-G(x,z)=\epsilon^{-1}\big(\tilde s(x,z;\widehat{\dot x})-s(x)\big)$ and, thus,
\begin{equation}
\label{eq:G_diff_bound}
\|\tilde G(x(t),z(t))-G(x(t),z(t))\|
\le
\epsilon^{-1} \ell_{s,e} \bar{e}(t_1)
\end{equation}
a.e. in $[t_0,t_1]$, with $\ell_{s,e}:=\sup_{x\in \cX}\|d(x)\nabla h(x)^\top\|$. 

Next, for the term $\|D s(x)\dot x\|$, since $s(\cdot)$ is Lipschitz on the compact set $\cX$ with constant $\ell_{s,x}$,
Rademacher's theorem implies $\|D s(x)\|\le \ell_{s,x}$ for a.e.\ $x\in\cX$.
Hence, for a.e.\ $t$, $\|D s(x(t))\,\dot x(t)\|\le \ell_{s,x}\,\|\dot x(t)\|$.
Moreover, using $\dot x = \sfN(x)+B(z-s(x))= \sfN(x)+B\tilde z$, one has
\begin{equation}
\label{eq:xdot_bound_pairing}
\|\dot x(t)\|
\le
\|\sfN(x(t))\|+\|B\|\,\|\tilde z(t)\|
\le
\bar N(t_1)+ \|B\|\,\|\tilde z(t)\|,
\end{equation}
where $\bar N(t_1):=\sup_{\sigma\in[t_0,t_1]}\|\sfN(x(\sigma))\|$.

Substituting~\eqref{eq:pairing_contraction}--\eqref{eq:G_diff_bound} into the bound on
$\llbracket \dot z;\tilde z\rrbracket$, and using~\eqref{eq:xdot_bound_pairing} one obtains, for a.e.\ $t\in[t_0,t_1]$,
\begin{align}
& \|\tilde z(t)\|\,D^+\|\tilde z(t)\| \le
-\epsilon^{-1}\|\tilde z(t)\|^2
+\epsilon^{-1}\ell_{s,e} \bar{e}(t_1)\|\tilde z(t)\| \nonumber \\
& ~~~~~~~~~~~~~~~~~~~~+\ell_{s,x}\big(\bar N(t_1)+\|B\|\,\|\tilde z(t)\|\big)\,\|\tilde z(t)\|.
\label{eq:pairing_scalar_pre}
\end{align}
If $\|\tilde z(t)\|>0$, divide both sides by $\|\tilde z(t)\|$; if $\|\tilde z(t)\|=0$, the resulting
inequality holds trivially by continuity of the right-hand side. Define $\lambda :=\epsilon^{-1} - \ell_{s,x}\|B\|$ and note that, since $\epsilon^{-1} > \ell_{s,x}\|B\|$, then $\lambda > 0$. Using this definition, one gets:
\begin{equation}
\label{eq:Dini_ey_final_pairing}
D^+ \|\tilde z(t)\|
\leq - \lambda \|\tilde z(t)\|
+\ell_{s,x}\bar N(t_1)
+\epsilon^{-1}\ell_{s,e} \bar{e}(t_1).
\end{equation}
Applying the Gr\"{o}nwall Comparison Lemma for absolutely continuous functions on $[t_0,t_1]$~\cite{desoer2009feedback} gives, for all $t\in[t_0,t_1]$,
\begin{align}
\|\tilde z(t)\| \le & e^{-\lambda(t-t_0)} \|\tilde z(t_0)\| \nonumber \\
&  \hspace{-.2cm}
+\frac{1-e^{-\lambda(t-t_0)}}{\lambda}
\Big(\ell_{s,x}\bar N(t_1)+\epsilon^{-1}\ell_{s,e}\bar e(t_1)\Big).
\end{align}
Using $\lambda=(1-\epsilon\ell_{s,x} \|B\|)/\epsilon$ and since $1-e^{-\lambda(t-t_0)} \leq 1$ yield the main bound. 

\subsection{Proof of Theorem~\ref{thm:traj_deviation}}

The main result and the proof are for given vector norms on $\real^n$ and $\real^m$, and the corresponding induced norms. Subscripts are omitted to make the notation lighter.  

Recall that $\tilde x(t):=x(t)-x_s(t)$. Then, $\dot{\tilde x}
=
\sF(x)-\sF(x_s)+B\big(z-s(x_s)\big)$ (noting that $w(t)$ cancels, since the two systems have the same disturbance). Add and subtract $s(x)$:
\[
\dot{\tilde x}
=
\sF(x)-\sF(x_s)
+B\big(z-s(x)\big)
+B\big(s(x)-s(x_s)\big).
\]
By~\cite[Thm.~3.7]{FB:24-CTDS}, and using the properties of the logarithmic norms $\mu(A + B) \leq \mu(A) + \mu(B)$ and $\mu(A) \leq \|A\|$ one gets for almost all $t\in[t_0,t_1]$,
\begin{align*}
D^+\|\tilde x(t)\|
&\le
\operatorname*{ess\,sup}_{\xi\in\operatorname{co}\{x(t),x_s(t)\}}
\mu\!\big(D\sF(\xi)\big)\,\|\tilde x(t)\| \\
& ~~~~~~~
+\|B\|\,\|\tilde z(t)\|
+\|B\|\,\|s(x(t))-s(x_s(t))\| \\
& \hspace{-.9cm}\le
c_\sF\,\|\tilde x(t)\|+\|B\| \|\tilde z(t)\|+\|B\|\,\|s(x(t))-s(x_s(t))\|
\end{align*}
where $\operatorname{co}\{x(t),x_s(t)\}$ is the convex hull of $x(t)$ and $x_s(t)$; i.e.,
$\operatorname{co}\{x(t),x_s(t)\}
:=\{\theta x(t)+(1-\theta)x_s(t)\mid \theta\in[0,1]\}$.

By definition of the sets $\cI$ and $\cA$, one has $\|s(x(t))-s(x_s(t))\|=0$ for $t\in\cI$; hence, for almost all $t\in[t_0,t_1]$,
\begin{equation}
\label{eq:delta_dini_ineq_simplified}
D^+\|\tilde x(t)\|
\le
\Big(c_\sF+\|B\|\,\ell_{s,x}\mathbf 1_{\cA}(t)\Big)\|\tilde x\|+\|B\| \|\tilde z(t)\|.
\end{equation}
Applying the  Gr\"{o}nwall Comparison Lemma~\cite[Appendix~E]{desoer2009feedback} (for locally integrable functions), one gets
\begin{equation}
\label{eq:delta_varconst_simplified}
\|\tilde x(t)\|\le
e^{\int_{t_0}^{t} a(\sigma)\,d\sigma}\,\|\tilde x(t_0)\|
+\|B\|\int_{t_0}^{t} e^{\int_{\tau}^{t} a(\sigma)\,d\sigma}\,\|\tilde z(\tau)\|\,d\tau,
\end{equation}
with $a(t):=c_\sF+\|B\|\,\ell_{s,x}\mathbf 1_{\cA}(t)$ for brevity. Using the definition of $T_{\cA}$ yields $\int_{\tau}^{t} a(\sigma)\,d\sigma
\le
c_\sF(t-\tau)+\|B\|\,\ell_{s,x}\,T_{\cA}(t)$,
and substituting this into~\eqref{eq:delta_varconst_simplified} gives
\begin{align}
\|\tilde x(t)\|\le & 
e^{c_\sF(t-t_0)+\|B\|\,\ell_{s,x}\,T_{\cA}(t)}\,\|\tilde x(t_0)\| \nonumber \\
& 
+\|B\|\,e^{\|B\|\,\ell_{s,x}\,T_{\cA}(t)}
\int_{t_0}^{t} e^{c_\sF(t-\tau)}\,\|\tilde z(\tau)\|\,d\tau. \label{eq:delta_with_cF_only_simplified}
\end{align}

Next, bound $\|\tilde z (\tau)\|$ using
$\|\tilde z (\tau)\|\le e^{-\lambda(\tau-t_0)}\|\tilde z (t_0)\|+E(\epsilon, t_1)$ as in~\eqref{eq:z_bound}, where $\lambda = \epsilon^{-1} - \ell_{s,x}\|B\| > 0$ by the assumption of Proposition~\ref{lem:z_tracks_s_bound}. The final bound~\eqref{eq:tilde_x_simplified_bound} is obtained by noticing that   since $c_\sF<0$,
\begin{align*}
\int_{t_0}^{t} e^{c_\sF(t-\tau)} e^{-\lambda(\tau-t_0)}\,d\tau
\le
\int_0^{t-t_0} e^{-\lambda u}\,du
=
\frac{1-e^{-\lambda (t-t_0)}}{\lambda}
\end{align*}
which can be bounded by $\frac{1}{\lambda}$; moreover, 
$ \int_{t_0}^{t} e^{c_\sF(t-\tau)}\,d\tau
=
\frac{e^{c_\sF (t-t_0)}-1}{c_\sF}
=
\frac{1-e^{c_\sF (t-t_0)}}{-c_\sF}$. This concludes the proof.

\section{Illustrative numerical experiments}

As a representative application,  consider a power transmission system with inverter-based resources (IBRs) operating in a grid-following configuration~\cite{dorfler2023control}. In particular, IBRs provide droop control and virtual inertia.
The transmission network consists of buses $\cN=\{1,\dots,N\}$, with synchronous generators at $\cG\subset\cN$ and IBRs at $\cR=\cN\setminus\cG$. 
Active power injections follow $p_n(\theta)=\sum_{j\in\cN_n}|V_n||V_j|b_{nj}\sin(\theta_n-\theta_j)$ assuming constant voltage magnitudes. 
Algebraic buses can be reduced to ODE models via frequency-divider reductions.

For generators $n\in\cG$, one can adopt the turbine--governor model~\cite{dorfler2023control}
\begin{subequations}
\begin{align}
\dot\theta_n &= \omega_n, \\
M_n\dot\omega_n &= -D_n\omega_n + p_{\mathrm m,n}-p_{\ell,n}-p_n(\theta), \\
\tau_n\dot p_{\mathrm m,n} &= -p_{\mathrm m,n}+p^{\mathrm r}_{n}-R_n\omega_n .
\end{align}
\end{subequations}
For grid-following IBRs, one approach is to use the model~\cite{guggilam2018optimizing} 
\begin{subequations}
\begin{align}
\dot\theta_n &= \omega_n, \\
M_n\dot\omega_n & = -D_n\omega_n + p^{\mathrm r}_{\mathrm{IBR},n}-p_{\ell,n}-p_n(\theta)
\end{align}
\end{subequations}
with virtual inertia $M_n$ and damping $D_n$.  The IBR local state is $x_n = (\theta_n, \omega_n)$. Stacking all the dynamics yields $\dot x = \sF(x) + w$, where the vector $w$ collects the loads $\{p_{\ell,n}\}$, which are unknown and time varying,  and the reference points $\{p^{\mathrm r}_{n}\}$ and $\{p^{\mathrm r}_{\mathrm{IBR},n}\}$,  while $ \sF(x)$ models the network coupling and the droop controllers. 

Each IBR enforces the local safety constraint $\omega_n\ge \underline\omega$. 
The corresponding closed-form CBF filter is 
$$
s_n(x)=\max\{0,- M_n e_{w_n}^\top (\sF(x)+w) - M_n \alpha_n (\omega_n-\underline{\omega})\},
$$
where $e_{w_n}$ is a vector with a 1 in the position of $\omega_n$ and zero otherwise.  
Augmenting the inverter dynamics with the ideal filter gives
\begin{align}
M_n\dot\omega_n = -D_n\omega_n + p^{\mathrm r}_{\mathrm{IBR},n}-p_{\ell,n}-p_n(\theta) + s_n(x),
\end{align}
while the proposed dynamic implementation is
\begin{align}
M_n\dot\omega_n &= -D_n\omega_n + p^{\mathrm r}_{\mathrm{IBR},n}-p_{\ell,n}-p_n(\theta) + z_n, \\
\epsilon \dot z_n &= - z_n + \max\{0,z_n - M_n \tilde{\dot \omega}_n - M_n \alpha_n (\omega_n - \underline{\omega})\}, \nonumber 
\end{align}
where $\tilde{\dot \omega}_n$ is a derivative estimate.

The IEEE-14 bus system using the DC power-flow approximation is simulated. 
Generators are placed at buses $2,3,6,8$, with IBRs at the remaining buses. 
The frequency nadir constraint is $\underline{\omega}=59.5$ Hz. 
Derivatives are estimated via dirty derivative as in  Remark~\ref{rem:dirty_derivative} with $\tau_d=0.01$, and trajectories are computed using forward Euler with step $10^{-3}$.

\begin{figure}[t]
\centering 
\includegraphics[width=1.0\columnwidth]{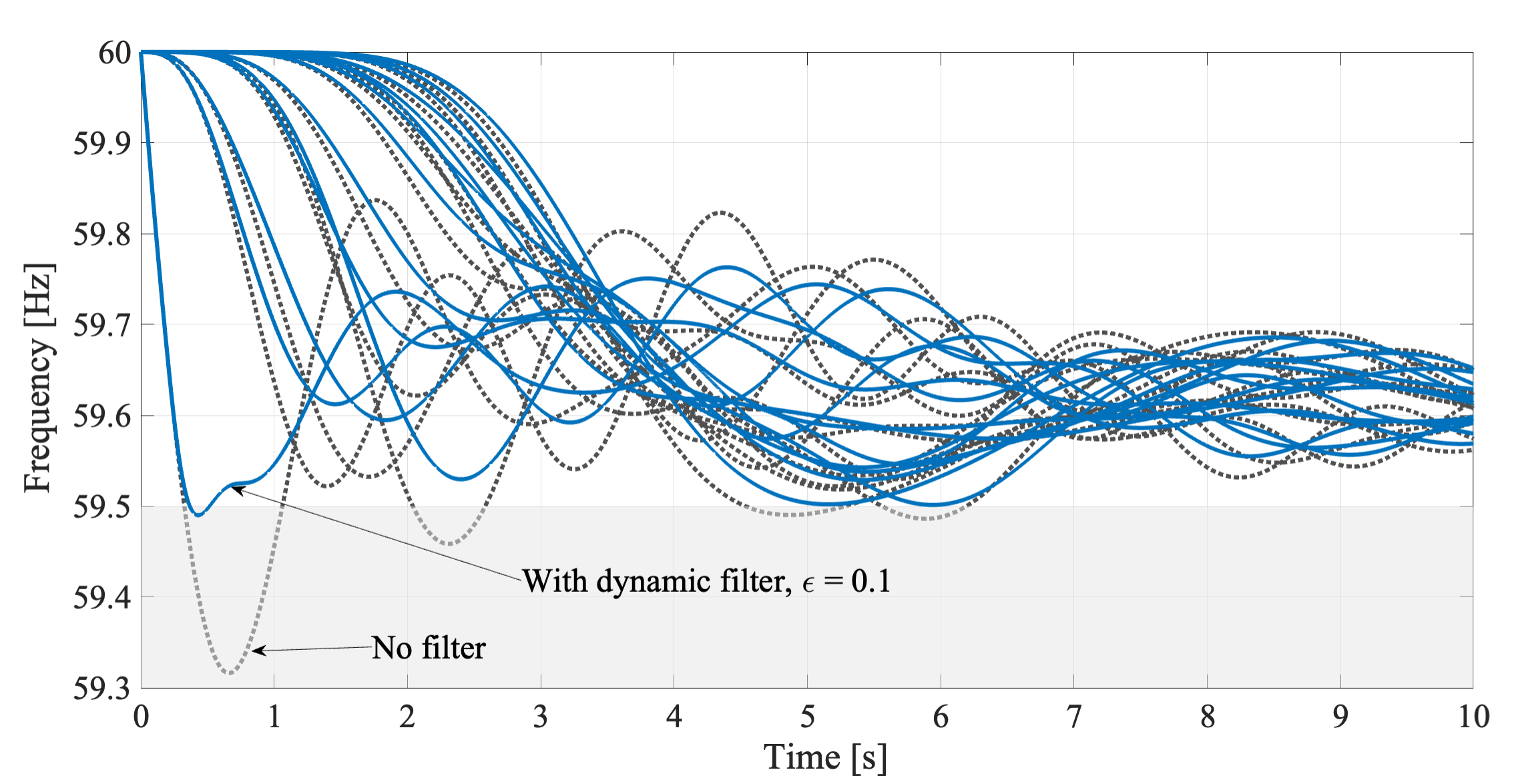}
\vspace{-.4cm}
\caption{Frequency across the system with and without dynamic filter.}
\vspace{-.2cm}
\label{fig:trajectories}
\end{figure}

\begin{figure}[t]
\centering 
\includegraphics[width=1.0\columnwidth]{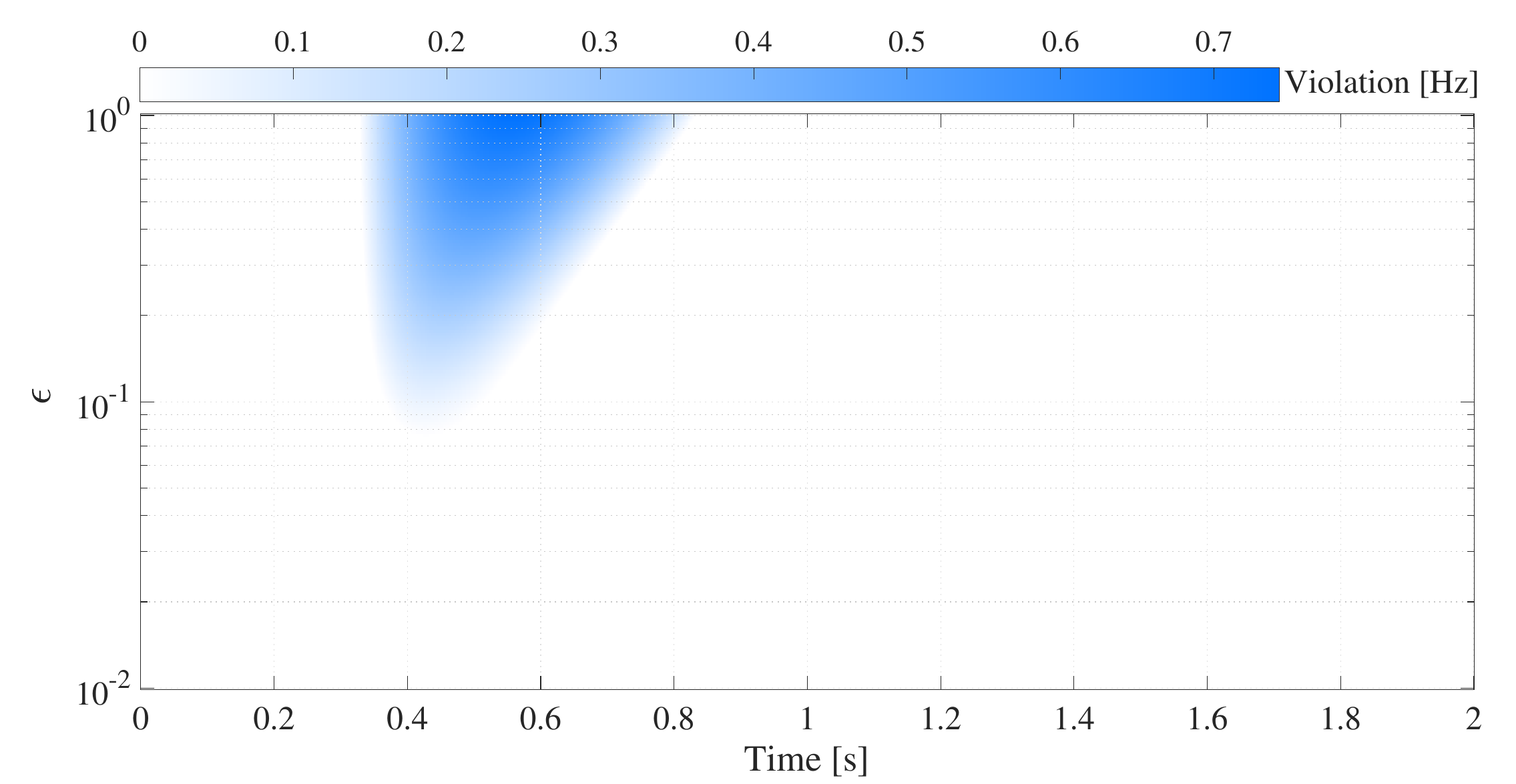}
\vspace{-.4cm}
\caption{Maximum lower-bound frequency violation as a function of time and $\epsilon$.}
\vspace{-.4cm}
\label{fig:violation}
\end{figure}

Figure~\ref{fig:trajectories} reports frequencies across all buses with and without the dynamic filter ($\epsilon=0.1$), showing effective post-disturbance regulation despite the fully local implementation. Figure~\ref{fig:violation} illustrates the magnitude of the frequency violation of the dynamic filter for different values of $\epsilon$ over the first few seconds after the disturbance. As $\epsilon$ decreases, the violation becomes smaller and is confined to a shorter time interval. For sufficiently small values of $\epsilon$, there is no violation of the frequency limit.

Connecting to Remark~\ref{rem:closed_form}, note that one can consider upper and lower bounds on $\omega_n$. The filter still admits a closed form solution, but it is omitted due to space limitations.


\section{Concluding remarks}

This paper studied locally implementable approximations of CBF-based safety filters for networked dynamical systems. 
The design was based on  a two-time-scale approximation and estimates of the state derivatives. The paper derived explicit trajectory deviation bounds that quantify the mismatch with ideal CBF-filtered dynamics and characterize how safety degradation depends on the time-scale parameter $\epsilon$, estimation errors, and filter activation. 
These results provide a principled framework for understanding trade-offs between safety guarantees and implementability. 
Future work will explore extensions to broader classes of safety constraints and robust ways to guarantee invariance with dynamic filters.

\bibliographystyle{IEEEtran}
\bibliography{biblio}

\appendix

\textbf{Additional simulation parameters}. The test system is the IEEE 14-bus transmission network. Synchronous generators are located at buses $\mathcal G=\{2,3,6,8\}$, while the remaining buses are modeled as grid-following inverter buses. The network line data are taken from the IEEE 14-bus test case available at \url{https://labs.ece.uw.edu/pstca/pf14/pg_tca14bus.htm}.

The bus-dependent parameters are the inertia coefficients \(M_i\), damping coefficients \(D_i\), turbine time constants \(\tau_i\) for generator buses, and droop coefficients \(R_i\). These are listed in Table~\ref{tab:busparams}.

\begin{table}[h]
\centering
\caption{Bus-dependent parameters used in the simulations.}
\label{tab:busparams}
\begin{tabular}{c c c c c}
\hline
Bus & \(M_i\) & \(D_i\) & \(\tau_i\) & \(R_i\)\\
\hline
1  & 2.5 & 1.2 & 0    & 0    \\
2  & 3.0 & 1.0 & 0.12 & 0.05 \\
3  & 2.8 & 1.5 & 0.12 & 0.07 \\
4  & 2.2 & 1.1 & 0    & 0    \\
5  & 3.5 & 1.3 & 0    & 0    \\
6  & 2.0 & 0.9 & 0.8  & 0.03 \\
7  & 2.7 & 1.4 & 0    & 0    \\
8  & 3.2 & 1.6 & 0.8  & 0.09 \\
9  & 2.9 & 1.2 & 0    & 0    \\
10 & 2.6 & 1.1 & 0    & 0    \\
11 & 2.4 & 1.0 & 0    & 0    \\
12 & 3.1 & 1.3 & 0    & 0    \\
13 & 2.3 & 1.2 & 0    & 0    \\
14 & 2.8 & 1.1 & 0    & 0    \\
\hline
\end{tabular}
\end{table}

For the dynamic implementation, the dirty-derivative estimator
\begin{align*}
\dot \xi_i & = -\frac{1}{\tau_d}\xi_i+\frac{1}{\tau_d}\omega_i, \\
\widehat{\dot \omega}_i & = \frac{1}{\tau_d}(\omega_i-\xi_i)
\end{align*}
is used with $\tau_d=0.01$. The CBF gains are set to $\alpha_n =10$. 

The trajectories are computed with a forward-Euler discretization using $dt=10^{-3}$ seconds, with initial conditions $x(0)=0$ and $z(0)=0$. A step disturbance of magnitude $3$ p.u. is applied at bus~1. For the heatmap in Fig.~\ref{fig:violation}, the parameter $\epsilon$ is swept over a logarithmically spaced grid in the interval $[10^{-2},1]$.

\balance

\end{document}